Laboratory Astrophysics and the State of Astronomy and Astrophysics

Submitted by the

American Astronomical Society Working Group on Laboratory Astrophysics
http://www.aas.org/labastro/

Nancy Brickhouse - Harvard-Smithsonian Center for Astrophysics
nbrickhouse@cfa.harvard.edu, 617-495-7438

John Cowan - University of Oklahoma
cowan@nhn.ou.edu, 405-325-3961

Paul Drake - University of Michigan
rpdrake@umich.edu, 734-763-4072

Steven Federman* - University of Toledo
steven.federman@utoledo.edu, 419-530-2652

Gary Ferland - University of Kentucky
gary@pa.uky.edu, 859-257-879

Adam Frank - University of Rochester
afrank@pas.rochester.edu, 585-275-1717

Wick Haxton - University of Washington
haxton@u.washington.edu, 206-685-2397

Eric Herbst - Ohio State University
herbst@mps.ohio-state.edu, 614-292-6951

Keith Olive - University of Minnesota
olive@physics.umn.edu, 612-624-7375

Farid Salama - NASA/Ames Research Center
Farid.Salama@mail.arc.nasa.gov, 650-604-3384

Daniel Wolf Savin* - Columbia University
savin@astro.columbia.edu, 1-212-854-4124,

Lucy Ziurys – University of Arizona
lziurys@as.arizona.edu, 520-621-6525

*Co-Editors

# 1. Introduction

Laboratory astrophysics and complementary theoretical calculations are the foundations of astronomy and astrophysics and will remain so into the foreseeable future. The impact of laboratory astrophysics ranges from the scientific conception stage for ground-based, airborne, and space-based observatories, all the way through to the scientific return of these projects and missions. It is our understanding of the under-lying physical processes and the measurements of critical physical parameters that allows us to address fundamental questions in astronomy and astrophysics. In this regard, laboratory astrophysics is much like detector and instrument development at NASA, NSF, and DOE. These efforts are necessary for the success of astronomical research being funded by the agencies. Without concomitant efforts in all three directions (observational facilities, detector/instrument development, and laboratory astrophysics) the future progress of astronomy and astrophysics is imperiled. In addition, new developments in experimental technologies have allowed laboratory studies to take on a new role as some questions which previously could only be studied theoretically can now be addressed directly in the lab. With this in mind we, the members of the AAS Working Group on Laboratory Astrophysics, have prepared this State of the Profession Position Paper on the laboratory astrophysics infrastructure needed to ensure the advancement of astronomy and astrophysics in the next decade.

The field of laboratory astrophysics comprises both theoretical and experimental studies of the underlying physics that produce the observed astrophysical processes. We have identified six areas of physics as relevant to astronomy and astrophysics. Astronomy is primarily an observational science detecting photons generated by atomic, molecular, and solid matter physics. Our understanding of the universe also relies on knowledge of the evolution of matter (nuclear and particle physics) and of the dynamical processes shaping it (plasma physics). Hence, our quest to understand the cosmos rests firmly on scientific knowledge in atomic, molecular, solid matter, nuclear, particle, and plasma physics. Chemistry is implicitly included here as part of molecular physics. Additionally, it is worth noting that there is not always a 1-to-1 correspondence between observational band-passes and the needed laboratory astrophysics. For example, standard UV/visible diagnostics for probing astrophysical environments are redshifted to longer wavelengths in high $z$ objects. Also, models of chemical processes involving photons at one wavelength are used to understand environments at other wavelengths.

This position paper is organized as follows: Section 2 presents a brief historical overview of the funding and infrastructure support for laboratory astrophysics. In Section 3 we highlight some of the key issues necessary to ensure a healthy and vital laboratory astrophysics community. Proposed actions to be taken and estimated levels of support needed are provided in Section 4. In Section 5 we give a brief summary of the current state of the field and in Section 6 list our recommendations to ensure a vibrant laboratory astrophysics community in the coming decade.

# 2. Historical Overview

In past decades, much of the laboratory astrophysics work required to move astronomy and astrophysics forward was funded by programs in atomic, molecular, solid matter, nuclear, particle, and plasma physics. The needs of astronomy and astrophysics were substantially synergistic with the directions of forefront, fundamental research in these fields. As a result, astronomy and astrophysics benefited from laboratory astrophysics research in these six areas without having to support them at a level anywhere close to that required to meet the actual need.

The last decade, however, has seen the funding reality change drastically. A number of programs that previously supported laboratory astrophysics research are no longer doing so, particularly in the critically important areas of atomic, molecular, and solid matter physics. The research currently supported by these programs has diverged from the needs of the astronomy and astrophysics community. Atomic physics has moved heavily into cold atoms, Bose-Einstein condensates (BEC), and quantum computation and cryptography (collectively known as photonics) as well as ultrafast lasers. Molecular physics has acquired a biological orientation and solid matter physics has moved over to nano-science**.** From a funding perspective, laboratory astrophysics now lies on the boundary between fields and as a result, its support is now insufficient to keep up with the demands of astronomy and astrophysics.

University support for laboratory astrophysics has also diminished drastically over the past decade as many faculty members have retired and departments have opted to move in new research directions. Areas related to photonics, ultrafast lasers, biology, environmental, or nano-technology are very much in vogue, as opposed to classic atomic, molecular, and solid matter physics. As a result, few new faculty members have been hired in laboratory astrophysics, thus threatening the future supply of researchers knowledgeable in this field. These faculty members are necessary not just to carry out the needed laboratory astrophysics research but also to train graduate and undergraduate students, i.e., the next generation. Student participation in research is critical for the future vitality of the field. The reduced numbers of faculty members and their associated laboratories has also led to a diminishing of the infrastructure in laboratory astrophysics. Most of the instrumentation associated with lab astrophysics research is not commercially available, and loss of personnel also results in a loss of technical expertise - a commodity that cannot easily be replaced.

### 3. Ensuring the future of laboratory astrophysics

*Increased and steady support for laboratory astrophysics among the various agencies is critical.* Current support for laboratory astrophysics comes from a small number of insufficiently funded programs. Robust funding programs are necessary to maintain the core competency of the community and to ensure the development of future generations of laboratory astrophysicists. If the funding for current programs is not increased, significant research capabilities that have required decades to develop will be lost. These research programs cannot be turned off and on at will and if stopped would require a large infusion of financial support and many decades to re-achieve previous capabilities. For instance, a new laboratory can cost up to several million dollars, much more than is needed to support, maintain, and enhance current facilities. The impending lack of sufficient and appropriate laboratory astrophysics groups and facilities will impact the scientific return from future astronomy and astrophysics projects, which typically have budgets that dwarf the level of support provided for laboratory astrophysics.

*Explicit support for laboratory astrophysics by missions and projects* is essential to maximize their astronomical and astrophysical scientific return. Current laboratory astrophysics funding is insufficient to produce all the critical data needed to ensure successful scientific return from missions and programs. Mission and project support of laboratory astrophysics through competitively run three-to-four year grants at a level comparable to those grants supporting core competency will make a significant impact on the production of the needed laboratory astrophysics data. The current support arising from one-year grants linked to observing cycles does not address the long-term nature of laboratory astrophysics research.

*Faculty development in laboratory astrophysics* is necessary to ensure the health and vitality of laboratory astrophysics on university teaching faculties. We urge that the various agencies offer awards for the creation of new tenure-track faculty positions within the intellectual disciplines that comprise laboratory astrophysics. This is a particularly important issue as start-up packages for laboratory astrophysics hires can be costly. The aim of these awards should be to integrate research topics in laboratory astrophysics into basic physics, astronomy, chemistry, electrical engineering, geosciences, biology, meteorology, computer science, and applied mathematics programs, and to develop laboratory astrophysics programs capable of training the next generation of leaders in this field.

*Establish fellowships and prizes in the area of laboratory astrophysics.* Recognition for research in laboratory astrophysics should be given greater visibility in the community. The creation of graduate and post-doctoral fellowships in laboratory astrophysics will help to train the next generation of researchers in this field. The establishment of sanctioned awards, analogous to those given in other research areas, would serve to raise the profile of laboratory astrophysics. Possible societies to consider for the creation of such awards include the American Astronomical Society (AAS), the American Chemical Society (ACS), the American Physical Society (APS), or similar professional societies.

*Strong Instrumentation, Technology, and Facilities Development Programs* in laboratory astrophysics are needed to support the development, construction, and maintenance of state-of-the-art laboratory astrophysics instrumentation and facilities. Such programs exist for detector and instrument development for observatories. No similar programs dedicated to laboratory astrophysics currently exist. Such programs are vital for ensuring not just that the capabilities of the laboratory astrophysics community remain current with present astronomy and astrophysics needs but that they also prepare for planned future astronomy and astrophysics needs. The time scale for developing new laboratories and technical capabilities is comparable to that for new projects and missions. Correspondingly, support for long term laboratory astrophysics development is critically needed. A range of instrumentation, technology, and development programs in the relevant agencies should be developed that would be able to respond to a variety of needs for infrastructure that promote basic research in laboratory astrophysics. The instrument and technology development component should provide funds for the design and construction of state-of-the-art as well as innovative instruments and technologies that will enable new laboratory astrophysics measurements or calculations. The facility development component should provide support for open facilities dedicated to laboratory astrophysics needs.

*Provide adequate funding for databases.* Critically evaluated data are needed by those analyzing astronomical measurements and modeling the associated environments. True understanding is only possible when collections of the highest quality laboratory astrophysics data are utilized. The relevant agencies and departments need to coordinate their efforts. Database compilation and the associated, vital critical evaluation, is a skill that is developed over decades in many cases. Long term commitment of funds is essential.

*Maintain a vibrant community of scientists conducting laboratory astrophysics.* The synergy between laboratory work in a university setting and in NASA and DOE laboratories must be fostered. A model for funding needs to include a continued level of baseline support, involving strengthened R&A programs, with occasional term-limited spikes for the topical needs of new missions and facilities. An example of the latter is the two-year grants in support of the Herschel mission. This is the only viable model to maintain core competency in the field, ensure rapid response to ongoing mission and project needs, and provide a healthy balance among energy bands and among disciplines. Finally, there should be coordination among the federal agencies (NASA, NSF, DOE) that fund all these activities.

*The NRC should conduct a study charged with identifying and detailing the specific laboratory astrophysics needed by the astronomy and astrophysics community.* Such a study should build on the finding of this Decadal Survey and follow those findings down to a level of detail which is beyond the scope of the Decadal Survey. Specific items to address include, but are not limited to, the following:

1. Within the six areas of laboratory astrophysics, what are the specific sub-areas needed by the astronomy and astrophysics community?

2. What is the number of university groups working in each distinct sub-area needed to meet current laboratory astrophysics needs in each sub-area and to train the future generations to insure continued capabilities in each sub-area?

3. How to encourage and retain faculty, in terms of ensuring the future supply of laboratory astrophysicists and maintaining/revitalizing infrastructure in the field?

4. How to foster graduate student participation and Ph.D. theses in these areas?

5. How to support the development and maintenance of laboratories and their unique instrumentation for ground-breaking research?

6. How to coordinate the activities of the agencies and departments that benefit from a robust effort in laboratory astrophysics?

7. How to combine interdisciplinary teams and/or centers focused on solving specific complex problems (e.g., NASA's Astrobiology Institute) while continuing to fund programs to support ground-breaking ideas of individual researchers that could potentially revolutionize aspects of astrophysics and increase the scientific return from observatories?

# 4. Proposed actions to be taken and estimated levels of support needed

At NASA, much of the support for laboratory astrophysics has come through programs such as the Astronomy and Physics Research and Analysis program (APRA). This and other programs historically have supported research in atomic, molecular, and solid matter laboratory astrophysics. Over the past four funding competition cycles the APRA program has received ~ 29 laboratory astrophysics proposals each year. Of these typically ~ 9 are selected for funding each year. Including new and continuing grants, the APRA program supports ~ 27 laboratory astrophysics at a level of approximately $3 million per year. Considering the vast scope of atoms, ions, molecules, and solids for which data are needed, and that the need for these data ranges from the far infrared to the x-ray regime, supporting only 9 new such projects each is woefully inadequate.

If the mandate of the APRA and similar programs remains unchanged, then a trebling of the funding level would make a significant positive impact on the vitality for atomic, molecular, and solid matter laboratory astrophysics. Such an increase would accomplish two goals. First, it would provide support for the many highly rated proposals that must currently be declined due to insufficient funds. Second, an increase in funding will initially increase the odds of receiving an award. This increase in the likelihood of obtaining funding will serve to draw new researchers into laboratory astrophysics, thereby growing our national capabilities in this area and laying the foundation for a vibrant and strong future.

If, however, the mandate of APRA and similar programs were extended to include plasma, nuclear, and particle laboratory astrophysics, then a dramatically larger increase would be needed. An expansion of the scope of the laboratory astrophysics covered by these programs without a corresponding increase in the level of funding would have a serious detrimental effect on the laboratory astrophysics capabilities of this nation. While new research in plasma, nuclear, and particle laboratory astrophysics would be initiated, fewer researchers in atomic, molecular, and solid matter laboratory astrophysics would be able to obtain funding. The only way to offset changes to the mandates of current programs is to make certain that the changes are accompanied by meaningful increases in the funding level available. For that reason, we propose that adding plasma, nuclear, and particle laboratory astrophysics to programs such as APRA should be accompanied by at least a doubling of the available funding. This is in addition to the proposed factor of 3 increase discussed in the preceding paragraph. Taken together, for example, these changes and increases would grow the laboratory astrophysics portion of the APRA program to a level of approximately $18 million per year.

At NSF, some of the recent support for laboratory astrophysics has come through the Division of Astronomical Sciences and the Division of Chemistry. The interests of the Division of Physics, particularly in atomic and molecular physics, have moved away from those areas relevant to the needs of the astronomy and astrophysics community. There is some indication that Chemistry is moving away from these areas as well. We also note that typical single investigator grants in the NSF division of astronomical sciences are far too small to support most laboratory astrophysics activities. We therefore urge the funding in laboratory astrophysics supported by the Divisions of Astronomical Sciences and Chemistry be not just maintained but even increased to reflect both the loss in support from the Division of Physics and the true cost of the research in laboratory astrophysics.

At DOE, programs in the Office of Basic Energy Science, Chemical Sciences Division previously funded laboratory astrophysics related research in atomic and molecular physics**.** The interests of this division have recently shifted to other areas unrelated to the needs of the astronomy and astrophysics community.  We urge that the DOE re-evaluate the mandates of its current atomic and molecular physics programs and establish a level of support for laboratory astrophysics appropriate for the astronomy and astrophysics needs at DOE.  DOE supports plasma research through the Office of Fusion Energy Sciences and through the National Nuclear Security Agency. To date, plasma laboratory astrophysics research has typically been possible only when there was close synergy with research relevant to fusion. Where such synergy exists, DOE should encourage it. However, some areas of research, such as studies of collisionless shocks relevant to supernova remnants, have not been feasible because of this limitation.  We urge DOE to remove this limitation and to open their programs and unique facilities to such studies.  Additionally, we urge DOE to recognize the natural overlap between many of the atomic and molecular data needs of the agency for fusion and those of astrophysics.  We therefore urge the DOE to establish a stronger programmatic link between the plasma/fusion and laboratory astrophysics commuities through the creation of new funding sources for such jointly relevant research.

The requests for proposals (RFPs) for all future ground-based, airborne, and space-based observatories should include an explicit request for a detailing of the laboratory astrophysics data needed to analyze the data from these projects and missions in order to maximize the scientific return.  It is no longer appropriate to think of these data needs as someone else's problem.  The responsibility for ensuring the needed data are there when the observatory comes on line should lie with the proposers of the observatory.  For this reason, we propose an appropriate fraction for the budget of each project or mission be allocated to relevant laboratory astrophysics studies.  Considering the hundreds of millions to billions spent on each observatory and the few millions that would be allocated to laboratory astrophysics, it is clear that the lever arm is long and the return on the support for laboratory astrophysics research will be greatly magnified.

Faculty development programs will create the infrastructure at universities necessary to train the next generation of laboratory astrophysicists.  We propose the creation of a program similar to the unrelated NSF Division of Atmospheric Sciences (ATM) Faculty Development program which was closed in 2004.  That program funded grants of up to $400,000 per year for up to 5 years.  We urge the creation of such programs at the NSF, NASA, and DOE at funding levels similar to those of the closed ATM program but adjusted for inflation.

Instrumentation and technology development programs specifically designed for laboratory astrophysics are vital in order for the capabilities of the laboratory astrophysics community to keep pace with the astronomy and astrophysics needs of this nation.  Support for a number of grants each year at a level per three year grant of up to several million dollars per grant would go a long way toward filling this vital need.  This program could be modeled on the existing NSF Division of Astronomical Sciences Advanced Technology and Instrumentation (ATI) program, the National Optical Astronomy Observatory (NOAO) Telescope Systems Instrumentation

Program (TSIP), or similar programs at NASA. The need for such programs specifically tailored for laboratory astrophysics is highlighted by the fact that none of these existing instrumentation and technology development programs have a history of supporting laboratory astrophysics. Because the laboratory astrophysics needs at NASA and the NSF do not always overlap, each agency should develop separate instrumentation and technology programs.

Facilities and databases require long term commitments extending beyond typical competitively run three year research grants. Additionally the support needed for facilities can often exceed that provided by instrumentation and technology development programs. For example, upcoming observatories will study the chemistry of the cosmos. This will require laboratory studies of cold molecules using facilities such as electrostatic ion storage rings. No such facilities currently exist in the United States. The construction capital costs for such a facility are estimated at approximately $15 million and the operating costs at approximately $2 million per year. A successful example of such a facility program is the Joint Institute for Nuclear Astrophysics (JINA) which integrates and utilizes the results of experimental and laboratory nuclear physics with broad and comprehensive theoretical and astrophysical studies. JINA is supported by the NSF Division of Physics at a level of ~ $1.7 million per year.

## 5. Summary

Ensuring the vitality and future of astronomy and astrophysics requires strong national support for a healthy and viable laboratory astrophysics program. Over the past 10 years there have been a series of NASA-sponsored workshops dedicated to the assessment of the state of laboratory astrophysics (NASA Laboratory Astrophysics Workshop 1998, 2002, 2006). These workshops have tracked the increasing direness of the situation. The reports and white papers resulting from these workshops have been published and are listed in the references section.

Laboratory astrophysics has reached a point where it is ceasing to be a viable, productive field, just at a time when advances in experimental technology are opening new vistas in its applicability. We urge the Decadal Survey to consider the laboratory astrophysics infrastructure needs for astronomy and astrophysics. We also urge the Survey Committee to follow the prioritized proposals they make all the way through to the underlying laboratory astrophysics data needed and to recommend sufficient funding to ensure the long term vitality of laboratory astrophysics and thereby of astronomy and astrophysics as a whole.

Without laboratory astrophysics, the scientific return from current and future NASA, NSF, and DOE observatories will diminish significantly. Funding for laboratory astrophysics has a long lever arm and any additional support will have significant impact on missions and facilities. Without laboratory astrophysics the future progress of astronomy and astrophysics will be greatly hindered.

## 6. Recommendations

- **Increased and steady support for laboratory astrophysics among the various agencies is critical.**

- **Explicit support for laboratory astrophysics by missions and projects is essential to maximize their astronomical and astrophysical scientific return.**

- **Faculty development in laboratory astrophysics is necessary to ensure the health and vitality of laboratory astrophysics on university teaching faculties.**

- **Establish fellowships and prizes in the area of laboratory astrophysics.**

- **Strong Instrumentation, Technology, and Facilities Development Programs in laboratory astrophysics are needed to support the development, construction, and maintenance of state-of-the-art laboratory astrophysics instrumentation and facilities.**

- **Provide adequate funding for critically evaluated databases which are needed by those analyzing astronomical measurements and modeling the associated environments.**

- **Maintain a vibrant community of scientists conducting laboratory astrophysics.**

- **The NRC should conduct a study charged with identifying and detailing the specific laboratory astrophysics needed by the astronomy and astrophysics community.**